\documentclass{article}
\usepackage{bm}
\usepackage{amsmath}
\usepackage{graphicx}

\newcommand{\grad}{\mathop{\rm grad}\nolimits}
\renewcommand{\div}{\mathop{\rm div}\nolimits}

\begin{document}
\title{Mathematical modeling of thermal stabilization of vertical wells on high performance computing systems} 
\author{Natalia V. Pavlova, Petr N. Vabishchevich,  Maria V. Vasilyeva }

\maketitle
\begin{abstract}
Temperature stabilization of oil and gas wells is used to ensure stability and prevent deformation of a subgrade estuary zone.
In this work, we consider the numerical simulation of thermal stabilization using vertical seasonal freezing columns.

 A mathematical model of such problems is described by a time-dependent temperature equation with phase transitions from water to ice. 
The resulting equation is a standard nonlinear parabolic equation.

Numerical implementation is based on the finite element method using the package Fenics. 
After standard purely implicit approximation in time and simple linearization, we obtain a system of linear algebraic equations.
Because the size of freezing columns are substantially less than the size of the modeled area, we obtain mesh refinement near columns. 
Due to this, we get a large system of equations which are solved using high performance computing systems.
\end{abstract}

\section{Mathematical model}
We consider a mathematical model that describes the distribution of temperature with phase transitions at a given temperature $T^*$ in a domain $\Omega = \Omega^- \cup \Omega^+ $. 
Here $\Omega^+(t)$ is a domain with liquid phase, where the temperature is above the phase transition temperature
\[
\Omega^+(t) = \{ \bm x | \bm x \in \Omega, \quad T(\bm x, t) > T^* \}
\]
and  $\Omega^-(t)$ stands for a domain with solid phase,
\[
\Omega^-(t) = \{ \bm x  | \bm x \in \Omega, \quad T(\bm x, t) < T^* \}.
\]
The phase transition occurs at a phase change boundary $S=S(t)$.

For simulation of heat transfer with phase transitions, the classical Stefan model is used \cite{vab}, \cite{vas}. 
This model describes the thermal processes accompanied by a phase change media, absorption and release of latent heat.
We use
\begin{equation}
\label{1}
\left( \alpha(\phi) + \rho^+ L \phi' \right) \left( 
\frac{\partial T}{\partial t}   - \left( \frac{k}{\mu}  \grad p, \grad T \right)
\right)
- \div \left(\lambda(\phi)  \grad T \right) = 0, 
\end{equation}
where
$L$ is thelatent heat of the phase transition,
$k$ stands for the absolute permeability of the porous medium,
$\mu$  is the water viscosity and  $p$ is the pressure in the soil.

For the coefficients of the equation, we have the following relations
\[
\begin{split}
&\alpha(\phi) = \rho^- c^- + \phi (\rho^+ c^+ - \rho^- c^-), \\
&\lambda(\phi)         = \lambda^- + \phi (\lambda^+ - \lambda^-), 
\end{split}
\]
and
\[
\phi = \left\{
\begin{matrix}
0, & \quad T < T^*, \\
1, & \quad T > T^*,
\end{matrix} \right. 
\]
where $\rho^+, c^+$ и $\rho^-, c^-$ are the density and specific heat capacity of the melt and frozen zone, respectively. 

Since we consider the process of heat propagation in porous media, then for the coefficients we have:
\[
c^- \rho^- = (1-m) c_{sc} \rho_{sc} + m c_i \rho_i,
\]
\[
c^+ \rho^+ = (1-m) c_{sc} \rho_{sc} + m c_w \rho_w,
\]
where $m$ is the porosity. 
Indexes $sc, w, i$ denote the skeleton of the porous medium, water and ice.
For the coefficients of thermal conductivity in the melt and frozen zone, we have similar relationships
\[
\lambda^- = (1-m) \lambda_{sc} + m \lambda_i,
\]
\[
\lambda^+ = (1-m) \lambda_{sc} + m \lambda_w.
\]

In practice, phase transformations do not occur instantaneously and can occur in a small temperature range $[T^* - \Delta, T^* + \Delta]$ \cite{vab}.
As the $\phi$-function, we can take $\phi_{\Delta}$:
\[
\begin{split}
&
\phi_{\Delta} = \left\{
\begin{matrix}
0, & T \leq T^*-\Delta, \\
\dfrac{\strut T - T^* + \Delta}{\strut 2 \Delta}, & \quad T^*-\Delta < T < T^*+\Delta, \\
1, & T \geq T^*+\Delta,
\end{matrix}
\right. 
\\
&\phi'_\Delta = \left\{
\begin{matrix}
0, & T \leq T^*-\Delta, \\
\dfrac{\strut 1}{\strut 2 \Delta}, & \quad T^*-\Delta < T < T^*+\Delta, \\
0, & T \geq T^*+\Delta.
\end{matrix}
\right. 
\end{split}
\]
Then we obtain the following equation for the temperature in the domain $\Omega$:
\begin{equation}\label{2}
\left( \alpha(\phi_\Delta) + \rho_l L \phi'_\Delta \right) \left( 
\frac{\partial T}{\partial t}   +  \bm u \grad T
\right)
- \div (\lambda(\phi_\Delta) \grad T) = 0.
\end{equation}
The resulting equation (\ref{2}) is a standard nonlinear parabolic equation.

The equation (\ref{2}) is suplemented with the initial and boundary conditions
\begin{equation}
\label{2.1}
\begin{split}
T(\bm x, 0) = T_0, &\quad \bm x \in \Omega, 
\\
T = T_c, &\quad \bm x \in \Gamma_D, 
\\
-k  \frac{\partial T}{\partial n}  = 0, &\quad \bm x \in  \Gamma / \Gamma_D.
\end{split}
\end{equation}
Here $\Gamma_D$ is  a place of contact with the freezing columns.

\section{Finite element realization}
The equation (\ref{2}) is approximated using a finite element method.
We multiply the temperature equation by a test function $v$, and integrate it using the Green formula
\begin{equation}\label{T.2.1}
\begin{split}
\int_{\Omega} \left(  \alpha(\phi_\Delta)  + \rho_l L \phi'_\Delta  \right)
&	\frac{\partial T}{\partial t}   \, v	 \, dx   \\
&+\int_{\Omega}  \left( \lambda(\phi_\Delta) \grad T, \grad v \right) dx  = 0, \quad  \forall v \in H_0^1(\Omega).
\end{split}
\end{equation}
Here $H^1(\Omega)$  is a Sobolev space, 
which consists of the functions $v$ such that $v^2$ and $|\nabla v|^2$ has a finite integral in the $\Omega$ and $H_0^1(\Omega) = \{ v \in H^1(\Omega) : v |_{\Gamma_D} = 0 \}$.

To approximate in time, we apply the standard fully implicit scheme and use the simplest linearization (from the previous time level)
\begin{equation}\label{T.2.2}
\begin{split}
\int_{\Omega} \left(  \alpha(\phi^n_\Delta)  + \rho_l L \phi'^n_\Delta  \right)
&\frac{T^{n+1} - T^{n}}{\tau}  \, v \, dx  \\
&+\int_{\Omega}  \left( \lambda(\phi^{n}_\Delta) \grad T^{n+1}, \grad v \right) dx  = 0.
\end{split}
\end{equation}
In such a way, we arrive at the following classical variational formulation of the problem: to find $T \in H^1(\Omega), \, \left[ T(\bm x, t) - T_c(\bm x, t) \right]|_{\Gamma_D} \in H_0^1(\Omega)$ such that
\begin{equation}\label{T.2.3}
\begin{split}
\frac{1}{\tau} \int_{\Omega} \left(  \alpha(\phi^n_\Delta)  + \rho_l L \phi'^n_\Delta  \right)
& T^{n+1}   \, v \, dx 
+\int_{\Omega}  \left( \lambda(\phi^{n}_\Delta) \grad T^{n+1}, \grad v \right) dx  \\
& = \frac{1}{\tau} \int_{\Omega} \left(  \alpha(\phi^n_\Delta)  + \rho_l L \phi'^n_\Delta  \right)
T^{n}  \, v \, dx , \quad  \forall v \in H^1(\Omega).
\end{split}
\end{equation}

The process of solving equation (\ref{T.2.3}) can be represented as follows.
While $t_{cur} < t_{max}$:
\begin{enumerate}
\item Calculate $t_{cur}$:  $t_{cur} = t_{cur} + \tau$;
\item Save a previous time lavel values $T_{prev} = T$;
\item Recalculate the ambient temperature $T_{air}$:
\[T_{air} =  41 sin((2\pi(t_{cur}/86400 + 250))/365) - 10.2;\]
\item If the temperature of the soil is less than $T_{air}$, then the freezing columns  turn on, else turn off;
\item The temperature at the new time level are solved by a linear solver;
\item Write the results to a file.
\end{enumerate}

Numerical implementation is performed using the Fenics package \cite{fenics}.
For results visualization, the values at each time level were recorded in vtk file format that was visualized using ParaView program.

\section{Numerical results}
\begin{figure}
\begin{center}
\includegraphics[scale=0.5]{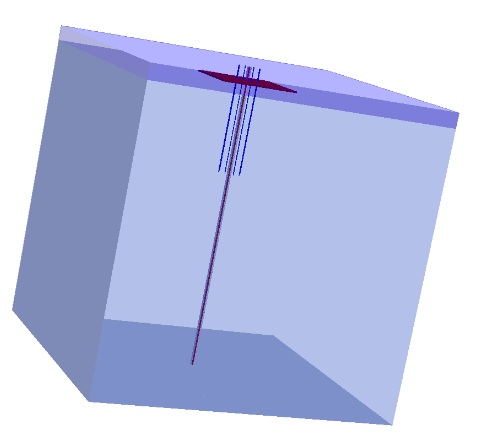} 
\end{center}
\caption{The computational domain}
\label{fig:netg}
\end{figure}

As a model problem, we consider the process of thermal stabilization of the mouth of the oil or gas wells \cite{snip}, \cite{fund1}.
The geometric domain was built using the Netgen mesher.
The computational domain is shown in Fig. \ref{fig:netg} and has a length of 40 meter in each direction, the well (radius 0.1 meter) is located in the middle of a field, where oil  flows with a given positive temperature.
A cement layer with the thickness of 0.2 meter is used for well heat insulation.
8 freezing columns with a radius of 0.05 meter are deepened to 14 meters around the well.
The top sand layer has the thickness of 2 meters. Near the well, there is laid penopleks (10 by 10 m, thickness of 200 mm).

Parameters of the problem are given in Table \ref{tab:T-2}.
The computational grid contains 10,903,946 cells.

\begin{figure}
\begin{center}
\includegraphics[scale=0.3]{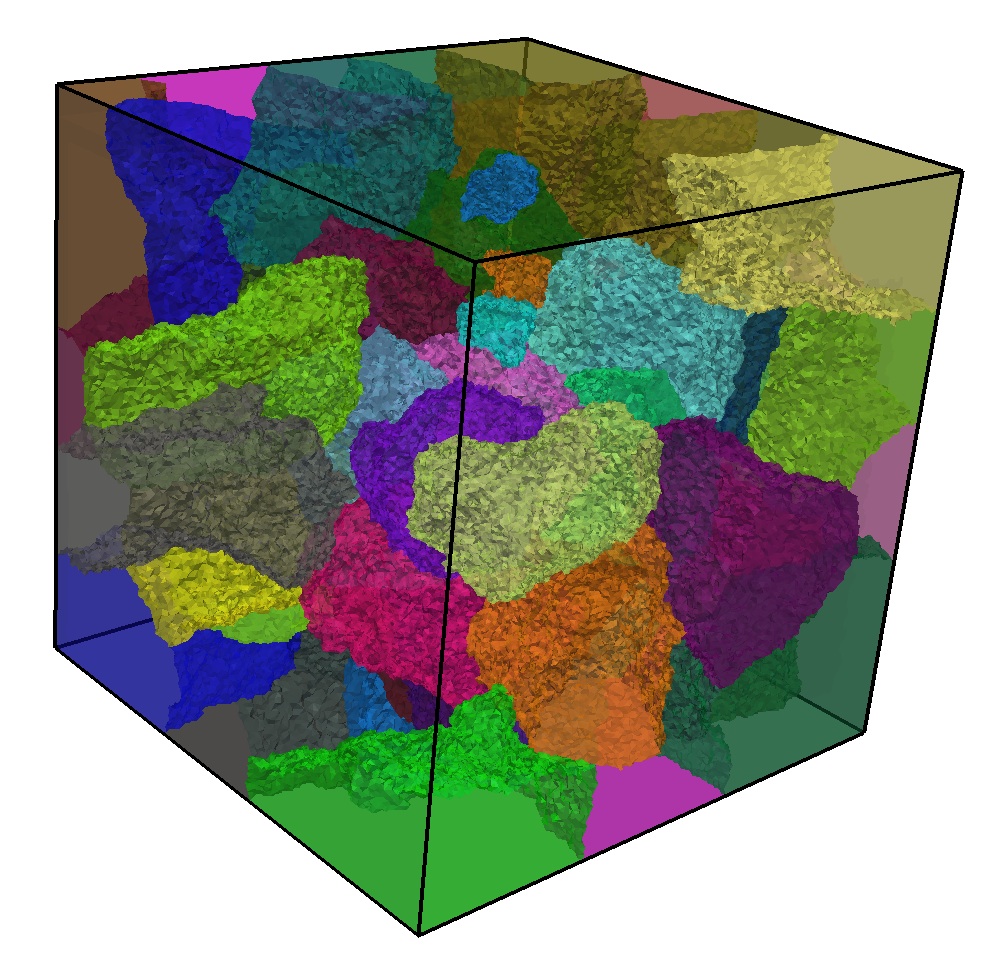} 
\end{center}
\caption{Domain decomposition}
\label{fig:parallel}
\end{figure}

\begin{figure}
\begin{center}
\includegraphics[scale=0.3]{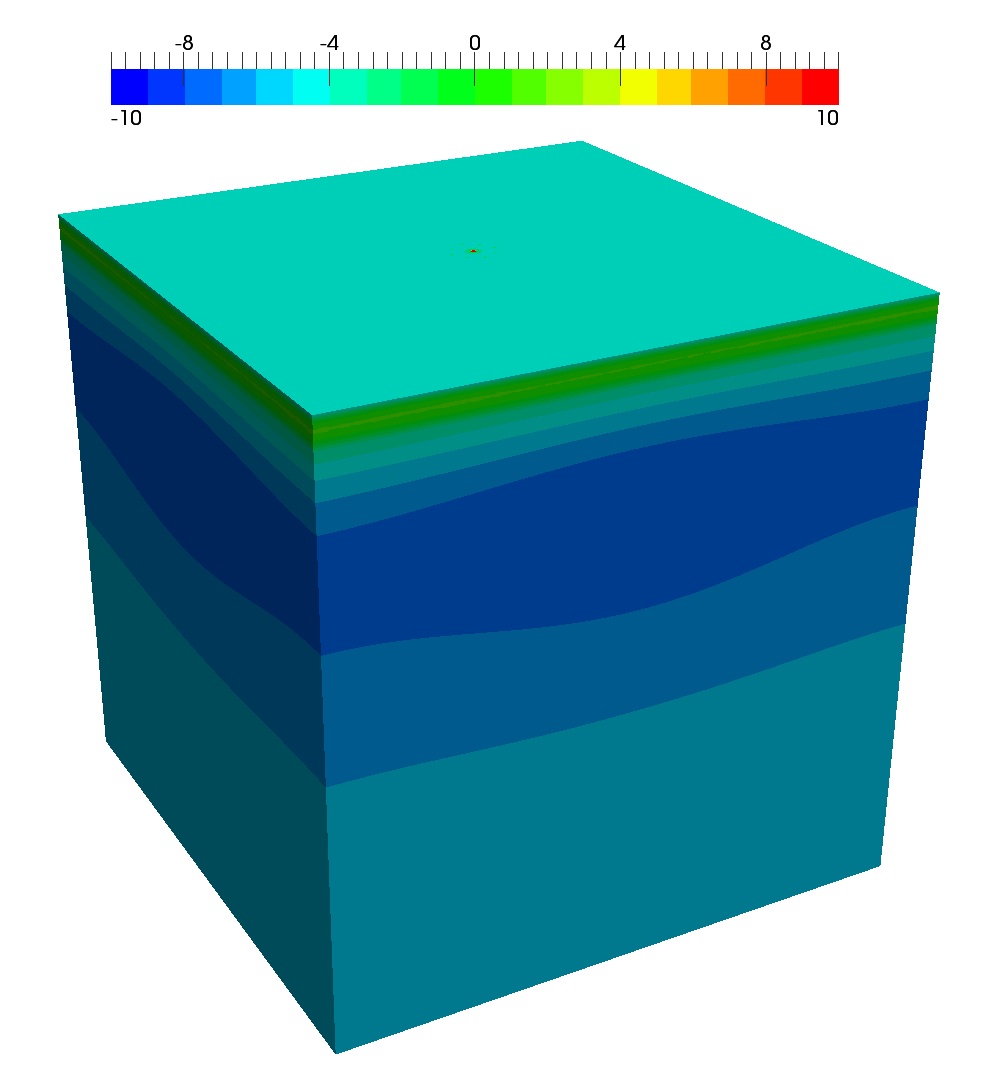} 
\end{center}
\caption{The temperature distribution after 5 years}
\label{fig:T5-3d}
\end{figure}

The numerical results are presented in Fig. \ref{fig:T5-3d}.  
Figure \ref{fig:T5-2d} illustrates the efficiency of the seasonal cooling devices, which accumulate the winter chill in the ground and provide additional bearing capacity in the summer.
By numerical simulation of the temperature stabilization of soils using freezing system, we can conclude that the presence of freezing columns can reduce soil thawing around wells
The calculations were performed using the NEFU computational cluster \textit{Ariane Kuzmin}.
The computation time using 64 processors (see 
Fig. \ref{fig:parallel}) was about 14 minutes, with 32 processors  - about 21 minutes, and with 16 processors - about 40 minutes, which shows good efficiency of parallelization.

\begin{table}
\begin{tabular}{llll}
\hline
Notation & Value & Metrics & Description \\
\hline 
$T_{cyl}$       &  20.0 &  degree & Temperature of oil in the well  \\
$T_0$           &  -5.0     &  degree &  Inital temperature \\
$T_*$             &  0.0     &  degree & Phase change temperature  \\
$L$               & 1.04e8  & J/kg  & Latent heat of the phase transition  \\
$c \rho_{sc}$           & 2.17e6  &  J/m3  &  Volumetric heat capacity of soil  \\
$c \rho_{l}$             & 2.42e6  &  J/m3  &  Volumetric heat capacity of water \\
$c \rho_{sa}$           & 1.34e6  &  J/m3  &  Volumetric heat capacity of sand \\
$c \rho_{pe}$          &  0.20e6   &  J/m3  & Volumetric heat capacity of polystyrene \\
$c \rho_{ce}$           &   0.8e6    &  J/m3  &  Volumetric heat capacity of cement \\
$\lambda_{sc}$        &  2.43   & W/(m degree)  &  The thermal conductivity of soil\\
$\lambda_{l}$          &  2.22 & w/(m degree)  &  The thermal conductivity of water \\
$\lambda_{sa}$        &  0.47 & W/(m degree)  &  The thermal conductivity of sand \\
$\lambda_{pe}$        &  0.03 & W/(m degree)  &  The thermal conductivity of polystyrene\\
$\lambda_{ce}$        &  0.21 & W/(m degree)  &  The thermal conductivity of cement \\
\hline 
\end{tabular} 
\caption{Problem parameters}
\label{tab:T-2}
\end{table}

\begin{figure}
\begin{center}
\includegraphics[scale=0.3]{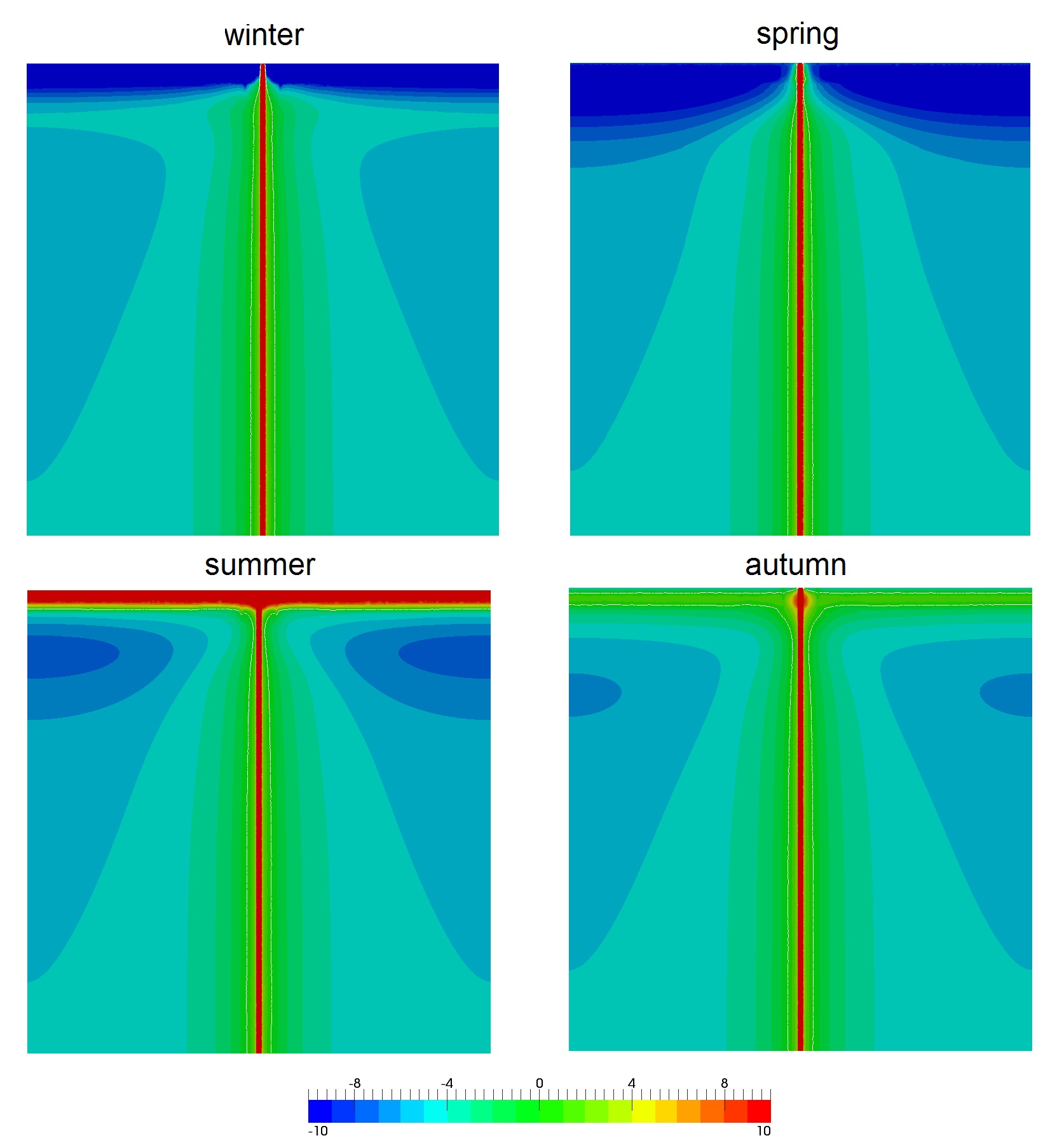} 
\end{center}
\caption{Temperature after five years. Slice $x = 20$. Without freezing columns}
\label{fig:T5-2d}
\end{figure}

\begin{figure}
\begin{center}
\includegraphics[scale=0.3]{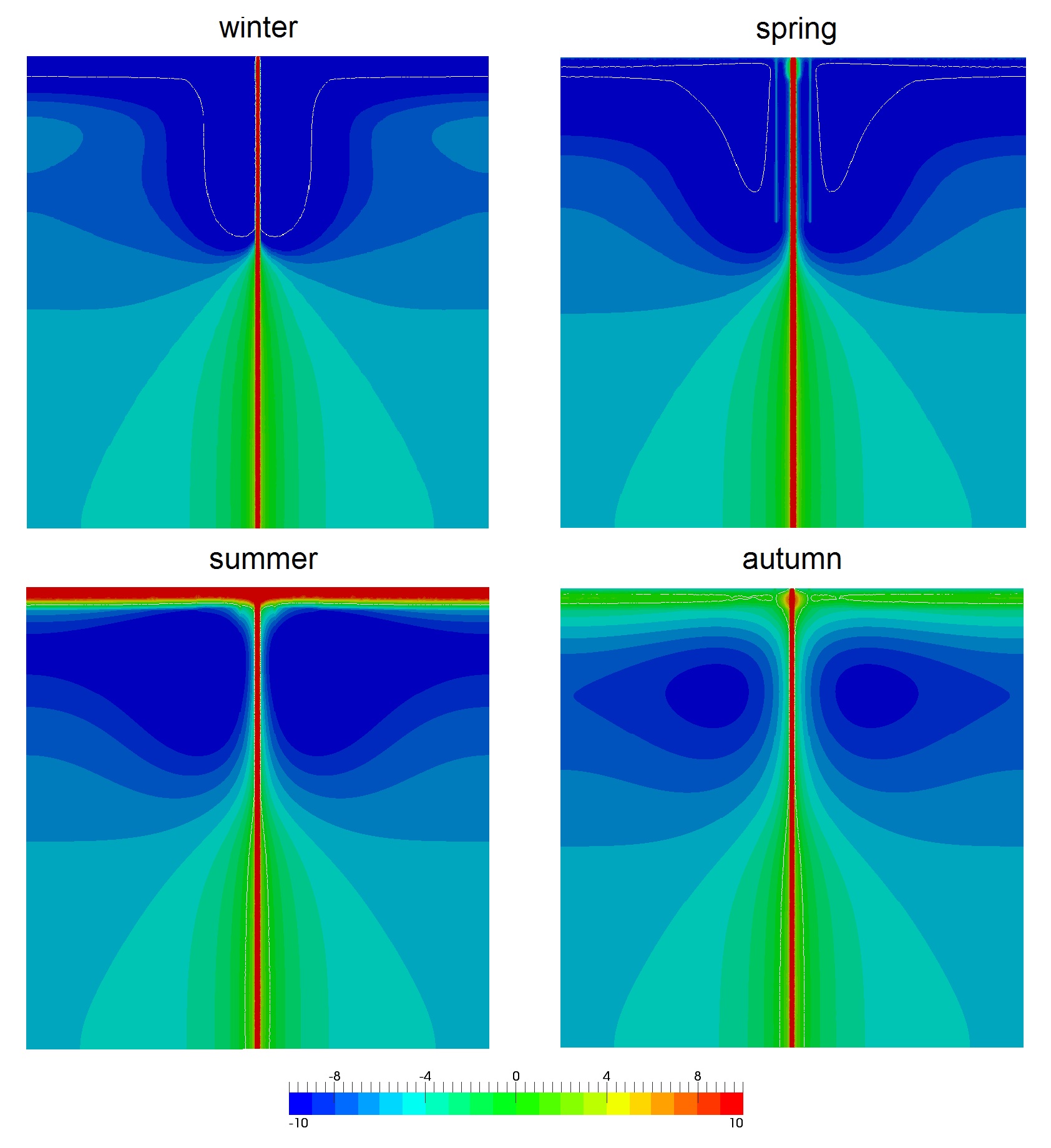} 
\end{center}
\caption{Temperature after five years. Slice $x = 20$. With freezing columns}
\label{fig:T5-2d}
\end{figure}

\bibliographystyle{splncs}

\begin{thebibliography}{99}  

\bibitem{vab}
Samarskii A. A.,  Vabishchevich P. N.: 
Computational Heat Transfer. 
Wiley, Chichester (1995)

\bibitem{vas}
Vasilyev V. I., Maksimov A. M., Petrov E. E., Cipkin G.G.: 
Heat and Mass Transfer in Freezing and Thawing Soils. 
Moscow, Nauka (1996)

\bibitem{snip}
SNIP 2.02.04-88
Foundations on permafrost.
State Construction Committee of Russia (2005)

\bibitem{fund1}
Tishchenko T.I., Gusev A. Y.: 
Technical solutions for the thermal stabilization of soils mouths of oil and gas wells.
Proceeding of the International Scientific-Practical Conference on Permafrost Engineering (2011)

\bibitem{fenics}
Logg A., Mardal K.-A., Wells G.:   
Automated Solution of Differential Equations by the Finite Element Method. 
Fenicsproject.org (2011)

\end{thebibliography}

\end{document}